\documentstyle[11pt,epsf,rotate]{article}

\begin{document}

\vspace*{1.5cm}


\begin{center}
{\Large\bf Interfering Doorway States and Giant Resonances.\\
 II.Transition Strengths}
\\[1cm]

V.V.~Sokolov$^{1}$, I.~Rotter$^{2,3}$,  D.V.~Savin$^{1}$ and M.~M\"uller$^4$
\\[8mm]
$^1$Budker Institute of Nuclear Physics, 6300090
Novosibirsk, Russia   \\
$^2$ Forschungszentrum Rossendorf, Institut f\"ur Kern- und Hadronenphysik,
\\D-01314 Dresden, Germany \\
$^3$ Technische Universit\"at Dresden, Institut f\"ur
Theoretische Physik,\\ D-01062 Dresden, Germany \\
 $^4$ Centro Internacional de Ciencias, Cuernavaca, Mexico\\

\end{center}

\vspace*{1cm}

\begin{abstract}
The mixing of the doorway components of a giant resonance (GR) due to the
interaction via common decay channels influences significantly the
distribution of the multipole strength and the energy spectrum of the decay
products of the GR.  The concept of the partial widths of a GR becomes
ambiguous when the mixing is strong.  In this case, the partial widths
determined in terms of the $K$- and $S$-matrices must be distinguished.  The
photoemission turns out to be most sensitive to the overlapping of the
doorway states. At high excitation energies, the interference between the
doorway states leads to a restructuring towards lower energies and  apparent
quenching of the dipole strength. \\
\end{abstract}

\newpage

\section[]{Introduction}
In \cite{one} we investigated analytically as well as numerically the dipole
giant resonance (GR) as a collective excitation in an open quantum system. In
the energy domain of the GR, both internal (due to the hermitian residual
interaction) and external (due to the interaction via common decay channels)
mixings are equally important. At the first stage, $k+1$ doorway states are
formed, with $k$ being the number of decay channels.  These states inherit
two different types of collectivity which are called, according to their
origin, internal and external collectivity, respectively.  The doorway
resonances formed in such a manner still interfere with one another due to
the external residual interaction. Finally a few resonance states with
appreciable dipole strengths are formed. The interference gives, generally,
rise to an essential redistribution of the dipole strength and shifts it
towards lower energies.

The investigations show further that two of the resonance states share the
main part of the total dipole strength and are therefore  most responsible
for the manifestations of the GR. The properties of these two doorway
components of the GR crucially depend on the degree of their overlapping. In
the case of weak overlapping they have comparable escape widths but the
dipole strength of the lower lying state is small. Quite opposite, a large
degree of overlapping leads to the appearence of two states with similar
dipole strengths whereas the escape width of the lower lying state is
dynamically reduced.

In the present paper, we study the cross section pattern in order to
elucidate the role of the external interaction and the interplay of both
types of collectivity in the experimentally measurable values.  Of special
interest are the transition strengths into specific channels when the
interaction via the energy continuum is strong.

In sect. 2, we describe the overlapping of doorway resonances in the context
of the general resonance scattering theory. The concept of the partial
escape widths in the case of overlapping resonances is reexamined from both,
inside ($K$-matrix) and outside ($T$-matrix) viewpoints. The transition
strengths in the particle channels are analytically analyzed in sect. 3. The
photoemission, which turns out to be especially sensitive to the degree of
overlapping of the doorway states, is studied in sect. 4. In sect.  5, we
discuss the interaction of the doorway states described in \cite{one} with
the background of complicated compound states which leads to an internal
damping of the collective exitation. We show in sect. 6 some numerical
results obtained in the same model (without internal damping), but with the
restrictions being removed which were introduced into the analytical
investigation. The numerical calculations confirm the main features of the
interference between the different types of doorway states as they follow
from the analytical study.  Finally, we summarize the results in sect. 7 and
draw some conclusions. Of interest is, above all, the apparent loss of the
collective dipole strength at high excitation energy.

All symbols used in this paper are the same as in \cite{one}.  We cite to an
equation in \cite{one} by writing its number in brackets with the upper
index [1], e.g. $(2.1)^{[1]}$ means eq. (2.1) in paper [1].  \\

\setcounter{equation}{0}
\section{Cross Sections and Partial Widths}
In the vicinity of an isolated resonance state $dw$ the hermitian
$K$-matrix is represented in the form
\begin{equation}\label{Rm}
\hat K(E) = \frac{\hat {\bf A}_{dw}^T\hat{\bf  A}_{dw}}{E - E_{dw}}
\end{equation}
where the row vector $\hat {\bf A}_{dw}$ is composed of the $k$ real decay
amplitudes $A_{dw}^c$ of the doorway state into the individual channels
$c=1,2,...,k$ and the superscript $T$ means transposition. The pole of this
matrix lies on the real energy axis at the energy $E_{dw}$ of the doorway
state. Eq. (\ref{Rm}) leads to the standard single-resonance Breit-Wigner
formula
\begin{equation}\label{BW}
{\hat T}_{dw}(E) = \frac{{\hat K}(E)}{1 + \frac{i}{2}{\hat K}(E)} =
\frac{\hat{\bf A}_{dw}^T\hat{\bf A}_{dw}}{E - E_{dw}
+ \frac{i}{2}\Gamma_{dw}}
\end{equation}
for the transition matrix. Though the pole of the transition matrix is
shifted to the point $E={\cal E}_{dw}=E_{dw}-\frac{i}{2}\Gamma_{dw}$ in the
complex energy plane, both matrices have the same residues. In particular,
the residues $\Gamma_{dw}^c=({A_{dw}^c})^2$ of the diagonal elements of these
matrices give the partial escape widths of the state $dw$ relative to the
channels $c$. The hermiticity of the $K$-matrix automatically provides the
unitarity of the scattering matrix $\hat S(E)=I-i\hat T(E)$ implying the well
known connection
\begin{equation}\label{Uc}
\Gamma_{dw} = \hat{\bf A}_{dw}^2
= \sum_c\Gamma_{dw}^c
\end{equation}
between the total width, $\Gamma_{dw}$, and the partial widths of the
resonance $dw$. In what follows we omit all nonresonant effects. They can,
if necessary, be easily taken into account by standard methods.

Using the parametrization (\ref{BW}), the partial widths of the resonance
state can be extracted from the experimental data. Averaging the cross
section of the reaction $c'\rightarrow c$ over all initial channels $c'$,
one obtains, with the help of the unitarity condition, the strength
\begin{equation}\label{Prb}
\sigma^c(E) = -\frac{\sigma_0}{\pi}\;{\rm Im}\,T_{dw}^{cc}(E) =
\sigma_0\;\frac{1}{2\pi}\frac{\Gamma_{dw}}
{\left(E - E_{dw}\right)^2 + \frac{1}{4}\Gamma_{dw}^2}\;\Gamma_{dw}^c =
\sigma_0\;\frac{2}{\pi}\,\frac{\Gamma_{dw}^c}{\Gamma_{dw}}\sin^2\delta(E)
\end{equation}
of the transition into the channel $c$.  Here, $\delta_{dw}(E)$ defined
by
\begin{equation}\label{delta}
{\rm tan}\,\delta_{dw}(E) = -\frac{1}{2}\,
\frac{\Gamma_{dw}}{E-E_{dw}}
\end{equation}
is the resonance scattering phase. The factor
$$\sigma_0\;\frac{1}{2\pi}\frac{\Gamma_{dw}}
{\left(E - E_{dw}\right)^2 + \frac{1}{4}\Gamma_{dw}^2}$$
describes the total cross section of the doorway state excitation.  Below
we set the factor $\sigma_0$ to unity measuring all cross sections in units
of this quantity. The maximal value
\begin{equation}\label{Br}
\sigma^c(E_{dw}) = \frac{2}{\pi}\;\frac{\Gamma_{dw}^c}{\Gamma_{dw}}
\equiv\frac{2}{\pi}\;{\cal B}_{dw}^c
\end{equation}
of the transition strength (\ref{Prb}) is proportional to the branching
ratio of the decay into the channel $c$.  The integration over the whole
resonance region gives the partial width itself,
\begin{equation}\label{Int}
\int_{-\infty}^{\infty} dE\;\sigma^c(E) = \Gamma_{dw}^c\;.
\end{equation}
Due to (\ref{Uc}) it follows from (\ref{Br}) and (\ref{Int}) that
\begin{equation}\label{Tot}
\frac{\pi}{2}\,\sum_c\sigma^c(E_{dw}) = 1\;, \quad
\sum_c\,\int_{-\infty}^{\infty} dE\;\sigma^c(E) = \Gamma_{dw}\;.
\end{equation}

The above discussion implies a good separation of the different resonance
states $dw$ so that any interference between them can be neglected. A more
careful analysis is however needed when the widths of the relevant doorway
states become comparable with their spacings. In this case one has to use
the formulae of the general theory of resonance reactions
\cite{F58,MW69,KobNO69,KirNO69}. Here, the transition matrix
\begin{equation}\label{Tm}
{\hat T}(E) = A^T\;\frac{1}{E - {\cal H}}\;A
\end{equation}
is composed of the three matrix factors which describe the formation of the
intermediate unstable system, its propogation and subsequent desintegration.
If there are $N_{dw}$ doorway resonance states near the excitation energy $E$
coupled to $k$ decay channels, the matrix $A$ consists of $k$
$N_{dw}$-dimensional column vectors ${\bf A}^c$ connecting all internal
states with each channel $c$. These vectors are real because of time-reversal
invariance.  In the following we neglect a possible smooth energy dependence
of the components $A_n^c$ over the whole energy domain considered.  The
validity of such an assumption is not always obvious and deserves a special
consideration. It may lead to further complications.

The evolution of the intermediate open system is described by the Green's
matrix
\begin{equation}\label{calG}
{\cal G}(E)=\frac{1}{E-{\cal H}}
\end{equation}
corresponding to the non-hermitian effective Hamiltonian
\begin{equation}\label{EffH}
{\cal H} = H - \frac{i}{2}AA^T = H - \frac{i}{2}\,W
\end{equation}
which has been investigated in detail in part I of this paper \cite{one}.
The factorized form of the interaction $W$ via the continuum ensures the
unitarity of the scattering matrix for arbitrarily overlapping resonances
\cite{KobNO69,KirNO69}. However, the simple Breit-Wigner parametrization
(\ref{BW}) loses its validity in general.

The propagator ${\cal G}(E)$ of the unstable system satisfies the Dyson
equation
\begin{equation}\label{Deq}
{\cal G}(E) = G(E) - \frac{i}{2}\;G(E)\;W\;{\cal G}(E)
\end{equation}
where
\begin{equation}\label{G}
G(E)=\frac{1}{E-H}
\end{equation}
is the resolvent of the hermitian part $H$ of the effective Hamiltonian
(\ref{EffH}). Subsequent iterations in the antihermitian part of the
effective Hamiltonian lead \cite{SZ89} to
\begin{equation}\label{Prp}
{\cal G}(E) = G(E) - \frac{i}{2}\;G(E)\;A\;\frac{1}{1 +
\frac{i}{2}{\hat K}(E)}\;A^T\;G(E)
\end{equation}
with
\begin{equation}\label{Rgen}
{\hat K}(E) = A^T\;\frac{1}{E - H}\;A = A^T\;G(E)\;A\;.
\end{equation}
The relation (\ref{Prp}) casts  again (compare with the first equality in
eq. (\ref{BW})) the transition matrix (\ref{Tm}) into the explicitly unitary
form
\begin{equation}\label{TmR}
{\hat T}(E) = \frac{{\hat K}(E)}{1+\frac{i}{2}{\hat K}(E)}\;.
\end{equation}

The  elements of both the $K$- and $T$- channel space matrices
\begin{equation}\label{TrfR}
K^{c\,c'}(E) = {\rm Tr}\,\left(G(E)\,{\bf A}^c\,({\bf A}^{c'})^T \right)\;,
\end{equation}
\begin{equation}\label{TrfT}
T^{c\,c'}(E) = {\rm Tr}\,\left({\cal G}(E)\,
{\bf A}^c\,({\bf A}^{c'})^T \right)\;,
\end{equation}
being the traces in the Hilbert space of the internal motion, are
independent of the choice of a basis in this space. In the eigenbasis of
the intrinsic hermitian part $H$ of the effective Hamiltonian (\ref{EffH}),
the matrix $\hat K$ is presented as the sum
\begin{equation}\label{Rmd}
{\hat K}(E) = \sum_r\frac{{\hat {\bf A}}_r^T{\hat {\bf A}}_r}
{E-\varepsilon_r}
\end{equation}
of pole terms similar to the single-resonance expression (\ref{Rm}).  The row
vectors ${\hat{\bf A}_r}$ consist of the real components
\begin{equation}\label{A_r}
A_r^c = {\bf\Phi}^{(r)}\cdot{\bf A}^c
\end{equation}
where the eigenvector ${\bf\Phi}^{(r)}$ of the hermitian matrix $H$ belongs
to the eigenenergy $\varepsilon_r$. The positive residues
\begin{equation}\label{Rpw}
\Gamma_r^c = (A_r^c)^2
\end{equation}
at the poles of the diagonal elements of the matrix (\ref{Rmd}), which
characterize the coupling of the intrinsic state ${\bf \Phi}^{(r)}$ to the
channels $c$, are the partial escape widths discussed in part I, eq.
$(2.17)^{[1]}$.

Analogously, the pole (resonance) parametrization of the transition matrix
(\ref{Tm}),
\begin{equation}\label{Tmd}
{\hat T}(E) =
\sum_{dw}\frac{{\hat {\bf A}_{dw}^T}{\hat {\bf A}}_{dw}}{E-{\cal E}_{dw}}
\end{equation}
is achieved by diagonalizing the total effective Hamiltonian (\ref{EffH})
(rather than only the hermitian part $H$ as above) with the help of a
transformation $\Psi$ which is complex since the Hamiltonian ${\cal H}$ is
not hermitian. Its complex eigenvalues
\begin{equation}\label{CE}
{\cal E}_{dw} = E_{dw} - \frac{i}{2}\Gamma_{dw}
\end{equation}
determine the energies and total widths of the overlapping resonance states.
The decay amplitudes of these states are (compare with (\ref{A_r}))
\begin{equation}\label{A_dw}
A_{dw}^c = {\bf\Psi}^{(dw)}\cdot{\bf A}^c
\end{equation}
with $\Psi^{(dw)}$ being the eigenvectors of the effective Hamiltonian
${\cal H}$. Together with these eigenvectors, the residues at the resonance
poles are also complex. Therefore, the resonances are mixed  in the
transition amplitudes with nonzero relative phases. In particular, the
residues are equal to
\begin{equation}\label{Tres}
\left(A_{dw}^c\right)^2 = |A_{dw}^c|^2\;{\rm exp}(2i\phi_{dw}^c)
\end{equation}
in the elastic scattering amplitudes. Here, the resonance mixing phases
$\phi_{dw}^c$ are introduced.

 Unlike the case of an isolated resonance described by eqs. (\ref{Rm},
\ref{BW}), the residues of the $K$- and $T$-matrices at individual poles do
not coincide if the doorway resonance states overlap. One can find the
connection between the decay vectors ${\hat{\bf A}}_r$ and ${\hat{\bf
A}}_{dw}$ starting with the eigenvalue problem ${\cal H}\,\Psi^{(dw)}={\cal
E}_{dw}\,\Psi^{(dw)}$ presented in the intrinsic eigenbasis of the hermitian
part $H$. Simple transformations lead then to the matrix equation
\begin{equation}\label{ArAdw}
\left[I + \frac{i}{2}\,{\hat K}({\cal E}_{dw})\right]\,
{\hat {\bf A}_{dw}} = 0\;.
\end{equation}
The determinant $\det\,\left[I + \frac{i}{2}\,{\hat K}({\cal E}_{dw})\right]$
is equal to zero at any resonance pole ${\cal E}_{dw}$ of the $T$-matrix
(\ref{TmR}). Therefore, for each resonance $dw$ a nontrivial solution of the
homogeneous linear system (\ref{ArAdw}) exists.  The proper solutions
are finally fixed by the Bell-Steinberger relation (\ref{BS}) (see below).

The square moduli
\begin{equation}\label{Tpw}
\Gamma_{dw}^c = |A_{dw}^c|^2\equiv
\big|{\bf\Psi}^{(dw)}\cdot{\bf A}^c\big|^2
\end{equation}
are just the quantities which are usually interpreted as the partial widths
of the resonance state $dw$. In the case of overlapping resonances, these
widths differ from the partial widths (\ref{Rpw}) defined in terms of the
$K$-matrix. Therefore we conclude that one has to distinguish between the
$T$-matrix partial widths (TPW) (\ref{Tpw}) extracted from the $T$-matrix,
and the $K$-matrix partial widths (KPW) (\ref{Rpw}) drawn from the matrix
$\hat K$.

The transformation matrix $\Psi$ satisfies the matrix equation
\begin{equation}\label{Psi}
{\cal H}\Psi = \Psi {\cal E}
\end{equation}
where ${\cal E}$ is the diagonal matrix of resonance energies ${\cal E}_{dw}$.
This transformation is complex orthogonal \cite{SZ89},
\begin{equation}\label{OC}
\Psi^T\Psi = \Psi\Psi^T = 1\;.
\end{equation}
However, for the hermitian matrix
\begin{equation}\label{U}
U = \Psi^{\dag}\Psi
\end{equation}
the inequality $U\neq I$ holds so that the overlapping resonance states are
not orthogonal (for illustration see \cite{PGR96}). The matrix $U$
appears in the well-known Bell- relation \cite{BS65} (see also
a compact matrix version of this relation in \cite{SZ89})
\begin{equation}\label{BS}
{\hat {\bf A}}_{dw}^*\cdot{\hat {\bf A}}_{dw'} =
i\,U_{dw\,dw'}\left({\cal E}_{dw'} - {\cal E}_{dw}^*\right)\;.
\end{equation}
Its diagonal part gives the relation
\begin{equation}\label{BSd}
\Gamma_{dw} = \frac{1}{U_{dw}}|{\hat {\bf A}}_{dw}|^2 =
\frac{1}{U_{dw}}\sum_c |A_{dw}^c|^2
\end{equation}
between the total widths and TPW (\ref{Tpw}). Here
\begin{equation}\label{Ud}
U_{dw} = 1 + 2\sum_n\left({\rm Im}\Psi_n^{(dw)}\right)^2 > 1
\end{equation}
is the corresponding diagonal matrix element of the matrix $U$. Because of
eqs.(\ref{BSd}) and (\ref{Ud}), the inequality condition
\begin{equation}\label{Inc}
\Gamma_{dw} < \sum_{c}\Gamma_{dw}^c
\end{equation}
holds in contrast to the equality (\ref{Uc}) characteristic for an isolated
resonance.

As it follows from eq. (\ref{BSd}), the TPW can be formally renormalized as
\begin{equation}\label{rTpw}
\tilde{\Gamma}_{dw}^c = \frac{1}{U_{dw}}\;\Gamma_{dw}^c\;,
\end{equation}
\cite{SB94,PGR96} leading
to the equality
\begin{equation}\label{Equ}
\Gamma_{dw} = \sum_{c}{\tilde\Gamma}_{dw}^c
\end{equation}
also for overlapping resonances. It should be emphasized however that neither
the $\Gamma_{dw}^c$ nor the renormalized quantities $\tilde\Gamma_{dw}^c$
coincide with the KPW $\Gamma_r^c$ from eq. (\ref{Rpw}) in the case of
overlapping resonances. The only relation between them,
\begin{equation}\label{TwRw}
\left({\bf A}^c\right)^2 = \sum_r\Gamma_r^c =
\sum_{dw}\Gamma_{dw}^c\;{\rm exp}(2i\phi_{dw}^c)
\leq \sum_{dw}\Gamma_{dw}^c\;,
\end{equation}
follows from the completeness of the sets of the corresponding eigenvectors.
Similarly, the energies $\varepsilon_r$ differ from the energies $E_{dw}$
of the resonance eigenstates. In the second equality (\ref{TwRw}) additional
phase factors appear in the sum over the resonance states. The imaginary part
of this sum vanishes since the contributions of different resonances
perfectly compensate one another.

The condition (\ref{TwRw}) results in the integral sum rules
\begin{equation}\label{SRd}
\int_{-\infty}^{\infty}dE\;\sigma^c(E) =
-\frac{1}{\pi}\int_{-\infty}^{\infty}dE\;{\rm Im}\;T^{c\,c}(E) =
\left({\bf A}^c\right)^2 = \sum_r\Gamma_r^c
\end{equation}
and
\begin{equation}\label{SRtot}
\sum_c\,\int_{-\infty}^{\infty}dE\;\sigma^c(E) = {\rm Tr}\,W =
\sum_{dw}\Gamma_{dw}
\end{equation}
instead of eqs. (\ref{Int}) and (\ref{Tot})  which are valid for an
isolated resonance. The integration is extended here over the whole energy
region, occupied by the overlapping resonance states.  Eq. (\ref{SRd})
leads to the sum of the KPW $\Gamma_r^c$, (\ref{Rpw}), rather than to the
sum of the TPW $\Gamma_{dw}^c$, (\ref{Tpw}).  Therefore, one cannot learn
much on the latter or even on their sum $\sum_{dw}\Gamma_{dw}^c$ from the
integral (\ref{SRd}) despite the expectation sometimes being expressed in
the scientific literature. Still less information can be drawn from the
maxima of the total cross section since their heights and positions are
connected with the widths and energies of the overlapping resonances in a
very complicated way.  At last, eq.  (\ref{SRtot}) fixes only the sum
of the total widths of all resonances.

A useful generalization of the sum rule (\ref{SRd}) reads
\begin{equation}\label{SRnd}
-\frac{1}{\pi}\int_{-\infty}^{\infty}dE\;{\rm Im}\;T^{c\,c'}(E) =
{\bf A}^c\cdot{\bf A}^{c'} = X^{c\,c'}
\end{equation}
where the $k\times k$ matrix \cite{M68,SZ89}
\begin{equation}\label{X}
\hat X = A^TA
\end{equation}
of the scalar products of the real amplitude vectors ${\bf A}^c$ appears.

\setcounter{equation}{0}
\section{Transition Amplitudes and Partial Transition
Strengths}
Similar to sect. 3 in \cite{one} (see eqs. $(3.1)^{[1]}$ and
$(3.2)^{[1]}$), we introduce the enlarged transition matrix
\begin{equation}\label{enT}
{\hat {\cal T}}(E) = {\cal A}^T\,{\cal G}(E)\,{\cal A}\;,
\end{equation}
\begin{equation}\label{enA}
{\cal A} = \left({\bf A}^0\equiv \sqrt{2i}{\bf D}
\;\;{\bf A}^1\;\;.\;.\;.\;\;\;{\bf A}^k\right)\;,
\end{equation}
containing along its main diagonal the function
\begin{equation}\label{enT00}
{\cal T}^{0\,0}(E)\equiv 2i\,{\cal P}(E) =
2i\;{\bf D}^T\,{\cal G}(E)\,{\bf D}
\end{equation}
besides the $k\times k$ block $T(E)$, (\ref{Tm}). The function ${\cal P}(E)$
together with
\begin{equation}\label{P}
P(E) = {\bf D}^T\,G(E)\,{\bf D}
\end{equation}
from $(3.3)^{[1]}$
is closely connected to the photoemission (see sect. 4).

The Green's matrix ${\cal G}(E)$, (\ref{calG}), is therefore needed for the
description of the evolution of the intermediate unstable system excited in
reactions. In  \cite{one}, a special doorway basis has been introduced which
is adjusted to the strong coherent nonhermitian interaction
\begin{equation}\label{effint}
{\cal H}^{(int)} = {\bf D}\,{\bf D}^T - \frac{i}{2}\,W
\end{equation}
(eq. $(4.1)^{[1]}$) by which the GR is created. In this basis, the
$(k+1)\times (k+1)$ doorway block  of the total Green's matrix is the only
one which has to be calculated. The influence of the trapped states
\cite{one} is included in a self-energy matrix which contains the coupling
between the doorway and trapped states. It manifests itself, as mentioned in
\cite{one}, in the fine structure variations of the transition amplitudes in
the energy region of the unperturbed parental levels.  Neglecting this fine
structure, one reduces the problem to the calculation of the Green's matrix
${\cal G}^{(dw)}(E)$ of the doorway effective Hamiltonian
\begin{equation}\label{effHdw}
{\cal H}^{(dw)} = \left( \begin{array}{cc}
{\cal H}^{(coll)} & \chi^T \\
\chi & \tilde{\cal H}
\end{array} \right)
\end{equation}
(eq. $(4.23)^{[1]}$).

The upper $2\times 2$ block
\begin{equation}\label{effHcoll}
{\cal H}^{(coll)} = \left( \begin{array}{cc}
\varepsilon_0 + {\rm sin}^2\Theta\,{\bf D}^2 &
{\rm sin}\Theta{\rm cos}\Theta\,{\bf D}^2 \\
{\rm sin}\Theta{\rm cos}\Theta\,{\bf D}^2 &
\varepsilon_0 + {\rm cos}^2\Theta\,{\bf D}^2
\end{array} \right)
-\frac{i}{2}\langle\gamma\rangle\left( \begin{array}{cc}
0 & 0 \\ 0 & 1
\end{array} \right)
\end{equation}
in (\ref{effHdw}) contains only two states which are strongly mixed by the
competing internal and external interactions characterized by the parameters
${\bf D}^2$ and $\langle\gamma\rangle$ respectively. Here ${\bf D}$ is the
$N$-dimensional vector of the dipole matrix elements, $\langle\gamma\rangle$
is the mean value of the nonzero eigenvalues $\gamma^c$ of the external
interaction matrix $W$ (or, equivalently, of the eigenvalues of the matrix
$\hat X$) and $\Theta$ stands for the angle between the dipole vector ${\bf
D}$ and the $k$-dimensional Hilbert subspace spaned by the $k$ decay vector
${\bf A}^c$.  We mark this block by the subscript ({\it coll}) since only
its eigenstates possess internal collectivity when the coupling $\chi$ is
neglected.

The $(k-1)\times (k-1)$ block $\tilde{\cal H}$ describes the $k-1$ doorway
states with energies close to $\varepsilon_0$ and mean widths
$\langle\gamma\rangle$. Contrary to the states of the first group, these
states carry no internal collectivity.

The two groups of doorway states are coupled via the continuum by the
antihermitian interaction
\begin{equation}\label{Chi}
\chi = - \frac{i}{2}\,\left({\bf 0}\;{\bf w}\right)\;.
\end{equation}
which
 can be expected to be moderately weak. Its strength is characterized
by the dispersion $\Delta_{\gamma}$ of the eigenvalues $\gamma^c$
(see $(4.28)^{[1]}$).

Representing the doorway Green's matrix ${\cal G}^{(dw)}(E)$ in the block
form complementary to (\ref{effHdw}), one obtains the following expression
$${\cal G}^{(coll)}(E) = \frac{1}{E-{\cal H}^{(coll)}-{\cal Q}(E)} =$$
\medskip
\begin{equation}\label{calGcoll}
\frac{1}{\Lambda(E)}\left( \begin{array}{cc} E - \varepsilon_0 - {\rm
cos}^2\Theta\,{\bf D}^2 + \frac{i}{2}\,\omega(E) & {\rm sin}\Theta{\rm
cos}\Theta\,{\bf D}^2 \\ {\rm sin}\Theta{\rm cos}\Theta\,{\bf D}^2 &
E - \varepsilon_0 - {\rm sin}^2\Theta\,{\bf D}^2 \end{array} \right)
\end{equation}
for its upper collective block with the function $\Lambda(E)$ given by
\begin{equation}\label{appSeqdw}
\Lambda(E)\equiv \left(E-\varepsilon_0\right)\,
\left(E-\varepsilon_{coll}\right)
+\frac{i}{2}\omega(E)\,\left(E-\varepsilon_0-
{\rm sin}^2\Theta\,{\bf D}^2\right) = 0\;;
\end{equation}
\begin{equation}\label{om(E)}
\omega({\cal E}) = \langle\gamma\rangle - \frac{i}{2}\,q({\cal E})\;.
\end{equation}
This result extends the formula for the Green's function $(3.11)^{[1]}$  of
the internal collective vibration  in a closed system to the consideration
of decaying collective modes.

In the doorway picture just described the elements of the matrix
(\ref{enT}) are presented as
\begin{equation}\label{calPdw}
{\cal P}(E) = {\bf D}^T\,{\cal G}^{(coll)}(E)\,{\bf D} =
{\bf D}^2\,\frac{E - \varepsilon_0 + \frac{i}{2}
{\rm sin}^2\Theta\,\omega(E)}{\Lambda(E)}
\end{equation}
\smallskip
\begin{equation}\label{Tdw}
T^{cc'}(E) = T^{cc'}_{coll}(E) + {\tilde T}^{cc'}(E)
\end{equation}
where
\begin{equation}\label{collTdw}
T^{cc'}_{coll}(E) =
\left(A_1^c-\frac{i}{2}\,q^c(E)\right)
\left(A_1^{c'}-\frac{i}{2}\,q^{c'}(E)\right)\,
\frac{E - \varepsilon_0 - {\rm sin}^2\Theta\,{\bf D}^2}{\Lambda(E)}
\end{equation}
and
\begin{equation}\label{tilTdw}
{\tilde T}^{cc'}(E) = \sum_{\alpha}\frac{A_{\alpha}^c\,A_{\alpha}^{c'}}
{E-{\tilde{\cal E}}_{\alpha}}\;.
\end{equation}
The quantities $A^c_1\;,A^c_{\alpha}$ (eq. $(4.10)^{[1]}$) are the components
of the (real) decay vectors ${\bf A}^c$ in the doorway basis. It is worthy
noting that the collective parts of the transition amplitudes vanish at the
energy
\begin{equation}\label{E_v}
E_v = \varepsilon_0 + {\rm sin}^2\Theta\,{\bf D}^2\;.
\end{equation}

The amplitudes (\ref{tilTdw}), being sums of independent Breit-Wigner terms,
contain themselves no interference effects. Indeed, all $A_{\alpha}^c$,
which connect the states inside the lower block of the Hamiltonian
(\ref{effHdw}) to the continuum, are real and (as one can easily check with
the help of eqs. $(4.36)^{[1]} - (4.38)^{[1]})$
\begin{equation}\label{tilgam}
\sum_c\left(A_{\alpha}^c\right)^2 = \tilde\gamma^{\alpha}\;.
\end{equation}
All interference effects are included in the collective part (\ref{collTdw}).
In particular, the mixing of the two different groups of the doorway
resonances in (\ref{effHdw}) is described by the selfenergy function
\begin{equation}\label{Q11}
q(E)\equiv - 4 {\cal Q}_{11}(E) =  \sum_{\alpha}
\frac{{w^{(\alpha)}}^2}{E-\tilde{\cal E}_{\alpha}}
\end{equation}
(eq. $(4.40)^{[1]}$), and the functions
\begin{equation}\label{qalph}
q^c(E) = \sum_{\alpha}\frac{w^{(\alpha)}\,A_{\alpha}^c}{E-
{\tilde{\cal E}}_{\alpha}}\;.
\end{equation}
All these functions are complex because of the complex doorway energies
${\tilde{\cal E}}_{\alpha}$. Therefore, although the dependence on the
channel indices $c,c'$ in the collective part (\ref{collTdw}) has the
desirable factorized form, the factors are generally complex and energy
dependent. As a result, the locations of the maxima in the cross sections
are not connected, contrary to the case of isolated resonances, with the
positions and the residues of the poles of the $K$- or $T$- matrices in any
simple way. If however the collective resonances do not overlap too strongly
all the functions $q(E)$ vary slowly within the energy region of the maximum
arising from the giant resonance state and can approximately be considered
as some complex constants.

The residues of the elastic reaction amplitudes are expressed in terms of the
complex energies of the doorway resonances as
$$ResT^{cc}({\cal E}_{dw}) = $$
\begin{equation}\label{ResT}
\left(A_1^c-\frac{i}{2}\,q^c({\cal E}_{dw})\right)^2
\left[1+\frac{1}{4}\,{\rm sin}^2 2\Theta\,
\frac{{\bf D}^4}{\left({\cal E}_{dw}-\varepsilon_0-{\rm sin}^2\Theta\,
{\bf D}^2\right)^2}+\frac{1}{4}\,q'({\cal E}_{dw})\right]^{-1}.
\end{equation}
In contrast to the real residues (\ref{Rpw}) of the $K$-matrix, they are
complex and carry information, hidden in the quantities $q^c$, on the
transition vectors $\hat {\bf A}_r$ of all the overlapping resonance states.
The concept of the $T$-matrix partial widths of GR, generally, becomes
irrelevant when its doorway components strongly overlap.  The only
information on the partial widths which one can extract from the
experimentally observed transition strengths $\sigma^c(E)$ is the sum rule
(\ref{SRd}) for the KPW.

The above formulae simplify appreciably if one neglects the coupling ${\bf
w}$ between  the two doorway blocks in (\ref{effHdw}). In such an
approximation  only the two upper collective doorway states $dw=0, 1$,
described in detail in subsection 4.3 of \cite{one}, share the total dipole
strength and contribute in the GR. The energy dependence of the
corresponding collective part
\begin{equation}\label{sigcoll}
\sigma_{coll}^c(E) =
\frac{1}{2\pi}\left(A_1^c\right)^2\,\langle\gamma\rangle
\,\frac{(E - E_v)^2}{(E-\varepsilon_0)^2(E-\varepsilon_{coll})^2+
\frac{1}{4}\langle\gamma\rangle^2\,(E-E_v)^2}
\end{equation}
of the total strength
\begin{equation}\label{sigtot}
\sigma^c(E) = -\frac{1}{\pi}\,{\rm Im}T^{cc}(E) = \sigma^c_{coll}(E) +
{\tilde\sigma}^c(E)
\end{equation}
of the transition into a particular decay channel $c$ turns out to have the
same universal form as in the single--channel model of ref. \cite{SZ90}.
In this respect, the expression (\ref{sigcoll}) is analogous to the
universal Breit-Wigner formula (\ref{Prb}).  According to
$(4.13)^{[1]}$, the condition
\begin{equation}\label{summa}
\sum_c (A_1^c)^2 = \langle\gamma\rangle
\end{equation}
is satisfied (compare (\ref{Uc})).

The hermitian $K$-matrix reduces in the same approximation to \cite{one}
\begin{equation}\label{UpsRtl}
{\hat K}(E) =
\frac{{\hat {\bf A}_d}^T\,{\hat {\bf A}_d}}{E-\varepsilon_{coll}} +
\frac{{\hat X}_{\bot}}{E-\varepsilon_0}
\end{equation}
with
\begin{equation}\label{ResK}
A^c_d = \left({\bf d}\cdot {\bf A}^c\right)\;,\qquad
{\hat X}_{\bot} = {\hat X} - {\hat {\bf A}_d}^T\,{\hat {\bf A}_d}\,.
\end{equation}
The strengths (\ref{sigcoll}) reveal two equally high maxima
\begin{equation}\label{sigmax}
\sigma_{coll}^c(\varepsilon_0) = \sigma_{coll}^c(\varepsilon_{coll}) =
\frac{2}{\pi}\,\frac{\left(A_1^c\right)^2}{\langle\gamma\rangle}
\end{equation}
just at the poles of the $K$-matrix (\ref{UpsRtl}).  Taking eq.
(\ref{summa}) into account, these relations are quite similar to eq.
(\ref{Br}).  Further, in close analogy with the first equation in
(\ref{Tot}),
\begin{equation}\label{Tot1}
\frac{\pi}{2}\,\sum_c\sigma_{coll}^c(\varepsilon_0) =
\frac{\pi}{2}\,\sum_c\sigma_{coll}^c(\varepsilon_{coll}) = 1\;.
\end{equation}
Nevertheless, at arbitrary values of the overlapping parameter $\lambda$,
the quantities $\left(A^c_1\right)^2$ coincide neither with KPW nor with TPW.
They are not the residues at the poles of the $K$- or $T$-matrices and
therefore cannot be ascribed to any internal eigenstates.

It could seem that the situation is improved by writing for example
\begin{equation}\label{Br1}
\sigma_{coll}^c(\varepsilon_{coll}) =
\frac{2}{\pi}\,\frac{\left(A_d^c\right)^2}{\langle\gamma\rangle
{\rm cos}^2\Theta} =
\frac{2}{\pi}\,\frac{\Gamma_{coll}^c}{\langle\gamma\rangle
{\rm cos}^2\Theta}\;.
\end{equation}
Here, $\Gamma_{coll}^c\equiv\left(A_d^c\right)^2=\left(A_1^c\right)^2\, {\rm
cos}^2\Theta$ are the KPW of the intrinsic collective state with the energy
$\varepsilon_{coll}$ while in the denominator
the sum  of all the widths, (\ref{summa}), stands. The same is valid for
the KPW $\Gamma_0^c=\left(A_1^c\right)^2\,{\rm sin}^2\Theta$ of the
intrinsic eigenstate with the energy $\varepsilon_0$. Nevertheless, it
should be stressed that the r.h.s. in (\ref{Br1}) is not the standard
branching ratio since the denominator $\langle\gamma\rangle{\rm
cos}^2\Theta$ has generally nothing to do with the total width of the
corresponding doorway state \cite{one}. Only in the limit $\lambda\ll 1$ of
a very weak overlapping this condition is fulfilled and the maxima of the
collective transition strengths provide the ordinary branching ratios ${\cal
B}_{dw=0,1}^c$, eq. (\ref{Br}), of the isolated doorway states $dw=0, 1$
$(4.52)^{[1]}$.

 However, the ratios of the KPW are
\begin{equation}\label{KG/KG'}
\frac{\Gamma_{r}^c}{\Gamma_{r}^{c'}} =
\frac{\sigma^c_{coll}(\varepsilon_{r})}
{\sigma^{c'}_{coll}(\varepsilon_{r})}\,,\qquad r=0, coll
\end{equation}
 independently of the value of $\lambda$. Thus, we conclude that the
parameters of the $K$-matrix can be directly extracted  from the maxima
of the collective part of the transition strengths $\sigma^c$.

The transition strengths (\ref{sigcoll}) drop  to zero at the point
$E=E_v$, eq.  (\ref{E_v}) which lies in between the two maxima. The maxima
are therefore well separated and their widths on the half heights may be
introduced in the two-level approximation. They can be explicitly found from
(\ref{sigcoll}) to be
\begin{equation}\label{G01/2}
\Gamma_{0;1/2} = \frac{1}{2}\,\left[1 - \frac{1}{2}\,
\left(\sqrt{1+\frac{4}{\lambda^2}+\frac{4}{\lambda}\,{\rm cos}2\Theta} -
\sqrt{1+\frac{4}{\lambda^2}-\frac{4}{\lambda}\,{\rm cos}2\Theta}\right)
\right]\langle\gamma\rangle
\end{equation}
and
\begin{equation}\label{G11/2}
\Gamma_{1;1/2} = \frac{1}{2}\,\left[1 + \frac{1}{2}\,
\left(\sqrt{1+\frac{4}{\lambda^2}+\frac{4}{\lambda}\,{\rm cos}2\Theta} -
\sqrt{1+\frac{4}{\lambda^2}-\frac{4}{\lambda}\,{\rm cos}2\Theta}\right)
\right]\langle\gamma\rangle\;.
\end{equation}
Although the sum
\begin{equation}\label{G0+G1}
\Gamma_{0;1/2}+\Gamma_{1;1/2}=\Gamma_{dw=0}+\Gamma_{dw=1} =
\langle\gamma\rangle
\end{equation}
depends neither on $\lambda$, nor on $\Theta$, each of
the terms of the sum does depend on the degree of overlapping. Thus, the
ratios
\begin{equation}\label{Br2}
\frac{\left(A_d^c\right)^2}{\Gamma_{dw;1/2}}
\end{equation}
do not characterize individual resonance states and cannot be interpreted
as their branching ratios.

The same is valid for the TPW. In the two-level approximation, the
residues (\ref{ResT}) at the poles ${\cal E}_{dw=0,1}$ can be presented in a
very simple form
\begin{equation}\label{appResT}
ResT^{cc}({\cal E}_{dw}) =
\frac{\left(A_1^c\right)^2}{\langle\gamma\rangle}\,
\Gamma_{dw}\,\frac{{\cal E}_{dw}-{\cal E}_{dw'}^*}
{{\cal E}_{dw}-{\cal E}_{dw'}}\;.
\end{equation}
This gives
\begin{equation}\label{collTpw}
\Gamma^c_{dw} =
\frac{\left(A_1^c\right)^2}{\langle\gamma\rangle}\,\Gamma_{dw}
\sqrt{\frac{1+
\left[{\rm tan}\,\delta_{dw}(E_{dw'}) -
{\rm tan}\,\delta_{dw'}(E_{dw})\right]^2}
{1+\left[{\rm tan}\,\delta_{dw}(E_{dw'}) +
{\rm tan}\,\delta_{dw'}(E_{dw})\right]^2}}
\end{equation}
for the TPW of the collective states.
Here, $\delta_{dw}(E_{dw'})$ is the scattering phase (\ref{delta})
on the resonance $dw$ taken at the energy of the resonance $dw'$. These
phases vanish only when the resonances are well isolated.

The last factor on the r.h.s. of eq. (\ref{collTpw}) is just the diagonal
matrix element $U_{dw}$, eq. (\ref{Ud}), of the Bell-Steinberger
nonorthogonality matrix (\ref{U}).  Using  the results of subsection 4.3 of
\cite{one}, one can present the latter factor explicitly in terms of the
mixing parameters $\Theta$ and $\lambda$,
\begin{equation}\label{Uexp}
U_{dw=0,1} = \frac{1}{\sqrt{2}}\left[1+\frac{1+\frac{1}{4}\lambda^2}
{\sqrt{\left(1-\frac{1}{4}\lambda^2\right)^2+\lambda^2{\rm cos}^2 2\Theta}}
\right]^{\frac{1}{2}}.
\end{equation}
In both limiting cases, $\lambda\ll 2$ and $\lambda\gg 2$, this factor goes
to unity while it is maximal in the intermediate region of $\lambda\approx
2$. In particular, for $\lambda=2$
\begin{equation}\label{Ulmbd2}
U_{0,1} = \left\{ \begin{array}{rl}
\frac{1}{\sqrt{1-{\rm tan}^2\Theta}}\,;\quad & 0<\Theta<\frac{\pi}{4} \\
\frac{1}{\sqrt{1-{\rm cot}^2\Theta}}\,;\quad &
\frac{\pi}{4}<\Theta<\frac{\pi}{2}\;.
\end{array} \right.
\end{equation}
The quantity (\ref{Ulmbd2}) becomes infinite for $\Theta=\frac{\pi}{4}$ as
mentioned in \cite{one}.

The factor $U$ disappears from the ratios
\begin{equation}\label{TG/TG'}
\frac{\Gamma^c_{dw}}{\Gamma^{c'}_{dw}} =
\frac{\sigma^c_{coll}(\varepsilon_{r})}
{\sigma^{c'}_{coll}(\varepsilon_{r})} =
\frac{\Gamma_{r}^c}{\Gamma_{r}^{c'}}
\end{equation}
of the TPW while the sum of $\Gamma^c_{dw}$
\begin{equation}\label{smrlT}
\sum_c\Gamma^c_{dw} = \Gamma_{dw}\,U_{dw}
\end{equation}
depends, contrary to the sums of the KPW, on the degree of overlapping
via the Bell-Steinberger factor $U$.

It has been shown in ref. \cite{SB94}, that the energy spectrum of the decay
products of an arbitrary two-level unstable system can generally be
expressed in terms of the  resonance energies ${\cal E}_{0,1}$, the
$T$-matrix "partial widths"
${\tilde\Gamma}^c_{dw}$, eq. (\ref{rTpw}),
which are renormalized  due to overlapping,
 and one additional real mixing
parameter which satisfies a sum rule following from the Bell-Steinberger
relation (\ref{BS}). The situation is even simpler in our quasi
single-channel case (see the remark below eq. (\ref{sigcoll})) where the
latter parameter is easily found explicitly \cite{SB94} as a function of the
complex resonance energies.  The resulting expression is remarkably simple,
\begin{equation}\label{sigcolldel}
\sigma^c_{coll}(E) =
\frac{2}{\pi}\frac{{\tilde\Gamma}^c_{dw}}{\Gamma_{dw}}\,
{\rm sin}^2\left[\delta_0(E)+\delta_1(E)\right]\;.
\end{equation}
(Note that, due to eq. (\ref{collTpw}), the ratio ${\tilde\Gamma}^c_{dw}
/\Gamma_{dw}$ is really the same for both doorway states $dw=0, 1$.)
This yields  for the transition strengths at the energy of a doorway
resonance
\begin{equation}\label{sigmaxdel}
\sigma^c_{coll}(E_{dw}) =
\frac{2}{\pi}\frac{{\tilde\Gamma}^c_{dw}}{\Gamma_{dw}}
\,{\rm cos}^2\delta_{dw'}(E_{dw})
\end{equation}
instead of eq. (\ref{Br}) for an isolated resonance.
The transition strengths do not attain their maximal values at the resonance
energies when the
resonances overlap. For this reason
 we have, in particular,  for the first
sum rule in (\ref{Tot})
\begin{equation}\label{Ltot}
\frac{\pi}{2}\,\sum_c\,\sigma^c(E_{dw}) =
{\rm cos}^2\delta_{dw'}(E_{dw}) < 1\;.
\end{equation}

One can easily convince oneself that both phases $\delta_{dw'}(E_{dw})$ drop
to zero when $\lambda\ll 2$ and the resonances are isolated. However, in the
opposite case of $\lambda\gg 2$ only the phase $\delta_0(E_1)$ of the narrow
resonance is small. The other phase, $\delta_1(E_0)$, belonging to the level
with the large width $\sim\langle \gamma\rangle$ is close to $\pi/2$. The
cross section (\ref{sigcoll}) has a narrow dip at the energy $E=E_v$ of the
state $dw=0$. In the limit of very large $\lambda$ the narrow state
decouples and gets invisible in the particle cross sections. At the same
time, this state acquires a large dipole strength due to the external
interaction \cite{one} and brightly manifests itself in the photochannel.

\setcounter{equation}{0}
\section[]{Photoemission}
The process of photoemission by the collective states turns out to be most
sensitive to their interference. To take the electromagnetic radiation into
account, one has to add to the antihermitian part of the effective
Hamiltonian ${\cal H}$ the new term
\begin{equation}\label{W_el}
-\frac{i}{2}\,W_{el} = -\frac{i}{2}\,\alpha_{el}\,{\bf D}{\bf D}^T
\end{equation}
describing the radiation of the same multipolarity as the internal coupling
vector ${\bf D}$. Therefore, the corresponding external coupling amplitude
\begin{equation}\label{Ael}
{\bf A}^{(rad)} = \sqrt{\alpha_{el}}\;{\bf D}
\end{equation}
is proportional to this vector with the constant $\alpha_{el}$ characterizing
the strength of the electromagnetic interaction.

The elastic matrix element of the $K$-matrix in the photo-channel is equal to
\begin{equation}\label{gamK}
K^{\gamma}(E) = \left({\bf A}^{(rad)}\right)^T G(E)\,{\bf A}^{(rad)} =
\alpha_{el}\,P(E)
\end{equation}
(see eq. $(3.3, 3.5)^{[1]}$). The radiation KPW are therefore proportional
to the dipole strengths $f^r = \left({\bf d}\cdot{\bf\Phi}^{(r)}\right)^2$,
eq. $(3.15)^{[1]}$, of the intrinsic eigenstates ${\bf\Phi}^{(r)}$,
\begin{equation}\label{phKPW}
\Gamma_r^{(rad)} = \alpha_{el}\;{\bf D}^2\,f^r\;.
\end{equation}
Since, according to eq.
$(3.24)^{[1]}$,
\begin{equation}\label{apstr}
f^1 = 1-\kappa^2\;, \qquad f^r \sim \frac{\kappa^2}{N-1}\;\;\; (r\neq 1)\;,
\end{equation}
one can immediately see that, in the limit of small $\kappa$, the internal
collective state appropriates the main part of the total radiation width
$\alpha_{el}{\bf D}^2$.  When $\kappa\rightarrow0$, only the pole at the
energy $\varepsilon_{coll}$ survives in the radiation $K$-matrix element.

The photoemission from the GR depends however upon the dipole strengths
${\tilde f}^s$ of the unstable doorway states ${\bf \Psi}^{(s)}$, eqs.
$(4.29,\; 4.31,\; 4.32)^{[1]}$, rather than upon the intrinsic quantities
(\ref{apstr}). It is easy to see that the photoelastic scattering amplitude
is obtained from the function ${\cal P}(E)$, (\ref{enT00}\,, \ref{calPdw}),
by substituting ${\bf D}^2$ by $(1-\frac{i}{2}\,\alpha_{el})\,{\bf D}^2$
when calculating the collective Green's matrix (\ref{calGcoll}). In the
two-level approximation, this leads to the result
$$\sigma^{(rad)}(E) = \frac{1}{2\pi}\,\alpha_{el}\,{\bf
D}^2\,\langle\gamma\rangle \times$$
\begin{equation}\label{sigrad}
\frac{(E-\varepsilon_0)^2\left({\rm cos}^2\Theta+\alpha_{el}/\lambda\right)
+\frac{1}{4}\,\alpha_{el}{\bf D}^2\,\langle\gamma\rangle\,{\rm sin}^4\Theta}
{\left[(E-\varepsilon_0)(E-\varepsilon_{coll})-
\frac{1}{4}\,\alpha_{el}{\bf D}^2\,\langle\gamma\rangle\,{\rm sin}^2\Theta
\right]^2 + \frac{1}{4}\,\langle\gamma\rangle^2\left[(1+\alpha_{el}/\lambda)
(E-\varepsilon_0)-{\rm sin}^2\Theta\,{\bf D}^2\right]^2}\;.
\end{equation}

For small values of the parameter $\lambda$, the principal maximum of the
photoemission strength lies at the energy $\varepsilon_{coll}$. Near this
point the expression (\ref{sigrad}) reduces to the standard Breit-Wigner
cross section
\begin{equation}\label{Srad}
\sigma^{(rad)}(E) = \frac{\Gamma^{(tot)}_{gr}}
{\left(E - \varepsilon_{coll}\right)^2 +
\frac{1}{4}\left[\Gamma^{(tot)}_{gr}\right]^2}\;\Gamma_{gr}^{(rad)}
\end{equation}
with the radiation and total widths
\begin{equation}\label{lm<1rad}
\Gamma_{gr}^{(rad)} = \alpha_{el}\,{\bf D}^2\,,\qquad
\Gamma_{gr}^{(tot)} = \langle\gamma\rangle\,{\rm cos}^2\Theta+
\alpha_{el}\,{\bf D}^2
\end{equation}
respectively. The giant resonance is formed in this case by the sole doorway
state $dw=1$ with $f^1={\tilde f}^1$=1.
With growing $\lambda$, the radiation branching  ratio ${\cal B}^{(rad)}=
\Gamma_{gr}^{(rad)}/\Gamma_{gr}^{(tot)}$ decreases as long as $\lambda$
does not approach the critical value 2.

The picture changes noticeably for very large values of
$\lambda$ ($\gg 2$). The main maximum is displaced to the point $E=E_v$,
(\ref{E_v}), where the transition strengths into the particle channels have
 an interference dip due to the narrow collective state $dw=0$. The
energy dependence is of Breit-Wigner shape but the radiation and total
widths become equal to
\begin{equation}\label{lm>1rad}
\Gamma_{0}^{(rad)}
= \alpha_{el}\,{\bf D}^2\,{\rm sin}^2\Theta = \alpha_{el}\,{\bf
D}^2\,{\tilde f}^0\,,\qquad \Gamma_{0}^{(tot)} =
\frac{1}{\lambda^2}\,\langle\gamma\rangle\, {\rm sin}^2
2\Theta+\alpha_{el}\,{\bf D}^2\,{\rm sin}^2\Theta\;.
\end{equation}
The peak contains only the part sin$^2\Theta$ of the total radiation
transition strength. It is naturally ascribed to the collective state $dw=0$
which acquired the dipole strength ${\tilde f}^0={\rm sin}^2\Theta$, (see
eq. $(4.55)^{[1]}$), due to the interaction via continuum. The nucleon width
of this state diminishes and the radiation branching ratio ${\cal
B}^{(rad)}$ increases together with $\lambda$. Therefore, the radiation
appears as a narrow line near the centroid of the broad resonance $dw=1$
which is visible only in the particle channels. The radiation from this
broad collective state is suppressed and manifests itself only as a long
tail which stretchs towards higher energies.  The radiation from the narrow
state $dw=0$ becomes therefore the brightest manifestation of the giant
resonance in the photoemission.

In the most interesting intermediate domain of parameters
$\alpha_{el}\ll\lambda\ll 1/\alpha_{el}$ the photo\-emission strength is
\begin{equation}\label{intlmb}
\sigma^{(rad)}(E)\approx -\frac{\alpha_{el}}{2\pi}\,
{\rm Im}\,{\cal P}(E)\;.
\end{equation}
The interference of the radiation from the two resonances is strongest when
$\lambda\approx 2$. The frequency spectrum of the radiation is broad in this
case, its characteristic width is $\sim {\bf D}^2$ and the radiation
intensity  remains small even in its maximum. Generally, the shape of
the spectrum is not Lorenzian when $\lambda\approx 2$.

\setcounter{equation}{0}
\section[]{Spreading Width}

We now discuss the interaction of the collective modes with the sea of the
complicated background states. The spectrum of the background states is
extremely dense at high excitations so that statistical methods are the only
relevant ones to use in this case.  As in \cite{SZ90}, we suggest that the
doorway states couple effectively to $N_{bg}\gg N_{dw}$ compound states
which lie in the energy domain of the GR and have no direct access to the
continuum. We also assume that the coupling matrix elements $V_{dw\,bg}$ are
random Gaussian variables with zero mean value. Then, after averaging over
the background fluctuations, the doorway Green's function changes in the
limit $N_{bg}\rightarrow\infty$ as ${\cal G}^{(dw)}(E)\rightarrow {\cal
G}^{(dw)}(E-\Delta+\frac{i} {2}\Gamma^{\downarrow})$ \cite{SZ90} where
$\Delta$ and $\Gamma^{\downarrow}$ are the energy shift and spreading width
respectively.  Neglecting their possible slow energy dependence in the whole
domain of the GR, we can fully incorporate the hermitian shift $\Delta$
(which is in fact small due to statistical reasons) into the mean position
$\varepsilon_0$.  The only effect of the interaction with the background
states is then the additional shift of the poles of the transition
amplitudes along the imaginary direction in the complex energy plane. Note
that  under such conditions the integral sum rule (\ref{SRd}) survives the
transformations made.

We will not present here the rather cumbersome general expressions.
Confining ourselves for the sake of simplicity to the two-level
approximation, the shift considered does not influence the relation
established in subsection 4.3 of ref. \cite{one} between the energies shifts
and dipole strengths of the collective doorway states. We suggest further
that the displacement ${\bf D}^2$ is smaller than both the escape and
spreading widths. It can then easily be shown that the transition
strength corresponding to the particle emission in a channel $c$ acquires
the Breit-Wigner shape
\begin{equation}\label{pchnl}
\overline{\sigma^c(E)}=\frac{\left(A_1^c\right)^2}{2\pi}\;
\frac{\Gamma_{tot}}{\left(E-E_{centr}\right)^2+\frac{1}{4}\,
\Gamma_{tot}^2}
\end{equation}
with the centroid $E_{centr}=\varepsilon_0+\cos^2\Theta\,{\bf D}^2$
and the total width $\Gamma_{tot}=\langle\gamma\rangle+\Gamma^{\downarrow}$.
Let us remind that the condition (\ref{summa}) holds for the quantities
$\left(A_1^c\right)^2$.

The evolution of the  averaged $\gamma$-strength
$\overline{\sigma^{(rad)}(E)}$, when the escape width $\langle\gamma\rangle$
changes from values smaller than $\Gamma^{\downarrow}$ to larger ones, is
appreciably richer. The strength transforms smoothly from
\begin{equation}\label{A}
\overline{\sigma^{(rad)}(E)}=\frac{1}{2\pi}\,\alpha_{el}{\bf D}^2\,
\frac{\Gamma^{\downarrow}}{\left(E-\varepsilon_{coll}\right)^2+
\frac{1}{4}\,\left(\Gamma^{\downarrow}\right)^2}
\end{equation}
for $\langle\gamma\rangle\ll\Gamma^{\downarrow}$ to
\begin{equation}\label{B}
\overline{\sigma^{(rad)}(E)}=\frac{1}{2\pi}\,\alpha_{el}\sin^2\Theta\,
{\bf D}^2\,\frac{\Gamma^{\downarrow}} {\left(E-E_v\right)^2+
\frac{1}{4}\, \left(\Gamma^{\downarrow}\right)^2}
\end{equation}
in the opposite limit $\langle\gamma\rangle\gg\Gamma^{\downarrow}$.
In the intermediate region, the maximum monotonously decreases and moves
towards lower energies. The shape of the radiation spectrum is not
Lorentzian when both widths are of comparable value. It is worthy noting
that the width of the $\gamma$-spectrum is always determined mainly by the
spreading width.  The escape width $\langle\gamma\rangle$ drops out not only
from eq. (\ref{A}) but also from eq. (\ref{B}). This is due to the fact that
the radiating state $dw=0$ becomes almost trapped.

Eq. (\ref{B}) implies the loss of an appreciable part ($=\cos^2\Theta$) of
the  radiation strength if the total escape width of the GR noticeably
exceeds the spreading width. The contribution of the broader collective
state which is described by the right long tail in Fig. 3(d) (see next
section) is invisible in eq. (\ref{B}). It is well known that the spreading
width in fact strongly exceeds the total escape width of giant resonances at
moderate excitation energies. However, in very hot nuclei the opposite
condition seems to be fulfilled. According to experimental data
\cite{CST87,BCG90} as well as theoretical arguments of statistical nature
\cite{LBBZ95}, the spreading width saturates with the excitation energy
whereas the escape width continues to grow.

\setcounter{equation}{0}
\section[]{Numerical Results}

The behaviour of the dipole strengths, energies and widths of the
interfering resonance states is reflected in the cross section pattern as
shown above analytically by using mainly the two-level approximation.
Below, we show the results of numerical investigations performed under less
restrictive assumptions. The  (purely illustrative) calculations are
performed with the same 10 levels and 3 channels as in \cite{one}.  Damping
is not taken into account, i.e. the results are true only for
$\langle\gamma\rangle,\;{\bf D}^2\gg\Gamma^{\downarrow}$ (see the discussion
in sect. 5).

In Figs. 1 to 3 we show the energy dependence of the transition strengths
into particle and photo channels for the three values of the overlap
parameter $\lambda$ = 0.1, 2 and 5. As in the figures in \cite{one}, the
energy $E$ is measured in units of the total energy displacement ${\bf
D}^2$. Due to the strong interference, the pattern is noticeably different
in the different final channels. One nicely sees the shift of the maximum at
the higher energy towards lower energies which is predicted by the two-level
approximation. Moreover, the fragmentation of the maximum at the lower
energy into a number of resonances can be seen which, of course, disappears
in the limit of degenerate unperturbed levels $e_n$. At last, the growing
restructuring of the dipole strength  with increasing external coupling in
favor of the lower-lying components is seen in Figs. 1(d) to 3(d). For
example the summed strength above $E>0$ amounts to 99\%, 87\% and 85\% in
the case of the degenerate unperturbed spectrum (dashed lines). As to the
maximum value of the transition strength into the photo channel at the
higher energy, it drops down by a factor of more than 10 when $\lambda$
increases from 0.1 to 2, while a narrow high peak appears in agreement with
the analytical consideration at lower energy when $\lambda$ becomes large.

The elastic and photo-nuclear reaction cross sections are shown in Figs. 4
and 5. They are calculated for the same three values $\lambda = 0.1, 2$ and
5 as the transition strengths in Figs. 1 to 3.  Both the shift of the
dipole resonance to lower energies and the loss of  its dipole strength are
seen very clearly also in these values.

Thus, the following scenario takes place. Provided that the coupling
(\ref{Chi}) is  negligible, the two  collective doorway states
$dw=0, 1$ fully exhaust the total dipole strength so that only they can
radiate $\gamma$-rays. The radiation pattern determined by these doorway
states turns out to be very sensitive to their degree of overlapping: as
long as the energy displacement of one of them is appreciably larger than
the sum of the particle escape widths (i.e. $\lambda\ll 1$) only one of them
radiates. If, however, they overlap ($\lambda\sim 1$) the interference leads
to a strong redistribution of the dipole strength as well as the escape
width between the two states. When the degree of overlapping exceeds some
critical  value $\sim 2$ the escape width of one of the states starts to
decrease (dynamical trapping effect). This effect is governed by the avoided
crossing of two resonances described in detail in \cite{mudiisro}.  In
the limit of strong overlapping, $\lambda \gg 1$, the nearly trapped state
acquires an appreciable dipole strength and therefore would radiate, in the
absence of any internal damping, a narrow electromagnetic line in the
vicinity of the centroid of the broad bump which is observed only in the
particle channels. The broad state, which also possesses noticeable dipole
strengths, contributes mostly to a long radiation tail stretched towards
larger energies.

The coupling (\ref{Chi}) admixes the other doorway states and leads to
an additional restructuring of the total dipole strength in favour of the
low-lying components.

\section[]{Summary}

On the basis of a phenomenological schematic model we investigated the
interferences between the doorway components of a giant multipole resonance.
The overlapping of different components influences significatly the
resonance spectrum and the cross section pattern since their interaction via
the energy continuum creates, at a certain critical value of the external
coupling, strong redistributions of the widths and dipole strengths of the
doorway states. The resulting GR pattern is formed mainly by two specific
collective doorway states. Both states possess
 comparable dipole strengths
but acquire essentially different escape widths.  While the broader state
determines the picture in the particle channels, the brightest feature in
the photoemission would be, in the absence of any internal damping, the
relatively narrow radiation line from another nearly "trapped" doorway
component which lies at somewhat lower energies.

The internal damping due to the coupling of the doorway states to the
background of complicated states smears out the effects of the interference
as long as the spreading width exceeds the total escape widths of the
doorway components. In very hot nuclei it is possible, however, that
the escape widths become larger than the spreading width \cite{CST87,BCG90}
which is expected to saturate with increasing excitation energy. If so, the
interference picture is not completely spoiled. The internal damping only
widens the line radiated by the narrow doorway state though it completely
masks the tail from the broad one. Therefore, the visible bulk of the GR
$\gamma$ emission originates from a specific state with dynamically reduced
particle escape width but large dipole moment (the trapped collective state)
while the emission from the broader state is suppressed being spread over a
wide energy range. This manifests itself as a seeming loss of a part of the
dipole strength of GR and as a shift of the GR to lower energy.

Both the shift down of a part of the dipole strength and the loss of some
part of the dipole strength itself are discussed at present in connection
with experimental results obtained for the excitation of collective modes in
hot nuclei (see e.g. the Proceedings of the Gull Lake Nuclear Physics
Conference on Giant resonances, 1993, \cite{gulllake}). The $\gamma$-ray
multiplicity from the decay of giant dipole resonances is shown
experimentally to increase with the excitation energy in agreement with the
100\% sum-rule strength as long as it is not too high.  At higher energies,
however, its saturation signals the quenching of the multiplicity and the
existence of a limiting
 energy for  the $\gamma$ emission
from the giant dipole resonance. The different existing theoretical
approaches can only partly explain the experimental situation observed
\cite{suo}.

The results obtained in the present paper point to a new mechanism which
could possibly shed an additional light on the problem. To our mind,
the saturation of the $\gamma$ multiplicity observed experimentally at about
250 MeV excitation energy in heavy nuclei \cite{gulllake,suo} may be, at
least partly, explained by the interference phenomena discussed in the
present paper. Further investigations of this interesting question are
necessary.

\vspace{1cm}

{\small
\noindent
{\bf Acknowledgment:}
We are grateful to E. Kolomeizew and  E. Persson for their interest in this
work.  One of us (V.V.S.) thanks V.G. Zelevinsky for discussion of the
results. The present investigations are supported by the Deut\-sche
Forschungsgemeinschaft (Ro 922/1,6), by the  Grant  94-2058 from INTAS
and by the Deutscher Akademischer Austauschdienst.


\vspace{1cm}

\section*{Figure Captions}
\vspace{.2cm}

\noindent {\bf Fig.1}      \\
The transition strengths into particle (a,b,c) and photo (d) channels for
$\lambda=0.1$ and the electromagnetic interaction strength $\alpha_{el}=0.01$.
The resonance states are the same as in Fig. 2
in \cite{one}. The dashed lines correspond
to the case of parental levels fully degenerated ($\Delta_e=0$). 
\\

\noindent {\bf Fig.2}        \\
The same as in Fig.1 but for $\lambda=2$.
\\

\noindent {\bf Fig.3}   \\
The same as in Fig.1 but for $\lambda=5$. Note the different $E$ scale in (d).
\\

\noindent {\bf Fig.4}     \\
The elastic cross section for $\lambda=0.1$ (a), $\lambda=2$ (b),
and $\lambda=5$ (c). The resonance states are the same as in Fig. 2
in \cite{one}. Note the different $E$ scale in (a).
\\

\noindent {\bf Fig.5}       \\
The photo-nuclear cross section for $\lambda=0.1$ (a), $\lambda=2$ (b),
and $\lambda=5$ (c). The resonance states are the same as in Fig. 2
in \cite{one}. Note the different $E$ scale in (c).
\\

\end{document}